\documentclass{amsart}

\usepackage[body={16cm,23cm}]{geometry}
\usepackage{cite}
\usepackage{amsmath,amssymb}
\usepackage{amsfonts}
\usepackage[mathscr]{eucal}
\usepackage{epic}
\usepackage{eepic}

\newcommand{\ds}{\displaystyle}
\newcommand{\uop}{\mathbf{u}}
\newcommand{\vop}{\mathbf{v}}

\newcommand{\xop}{\mathbf{x}}
\newcommand{\yop}{\mathbf{y}}

\newcommand{\alg}{\mathcal{H}}

\newcommand{\Nop}{\mathbf{h}}

\begin{document}

\vspace{2cm}

\dedicatory{Talk given at the conference ``New frontiers in
exactly solved models'', ANU, July 21-22, 2005}

\title[]{Integrability of $q$-oscillator lattice model}
\author{S. Sergeev}%
\address{Department of Theoretical Physics,
Research School of Physical Sciences and Engineering, Australian
National University, Canberra ACT 0200, Australia}
\email{sergey.sergeev@anu.edu.au}

\thanks{This work was supported by the Australian Research Council}%


\begin{abstract}
A simple formulation of an exactly integrable $q$-oscillator model
on two dimensional lattice (in $2+1$ dimensional space-time) is
given. Its interpretation in the terms of 2d quantum inverse
scattering method and nested Bethe Ansatz equations is discussed.
\end{abstract}

\maketitle

Let $Z$ be the partition function of the ice-type model
\cite{Baxter-book} on the rectangular lattice with the size
$N\times M$ and periodical boundary conditions. Let the lattice be
completely inhomogeneous, the ``weights'' of $j^{\,\textrm{th}}$
vertex ($j=1,...,NM$) are parameterized by
\begin{figure}[ht]
\setlength{\unitlength}{0.25mm}
\begin{picture}(400,100)
\put(0,50){\begin{picture}(50,50) \linethickness{0.1mm}
\put(0,25){\line(1,0){50}} \put(25,0){\line(0,1){50}}
\put(22,-30){\small $1$}
\end{picture}}
\put(70,50){\begin{picture}(50,50) \linethickness{0.5mm}
\put(0,26){\line(1,0){24}}\put(24,25){\line(0,1){24}}
\put(26,0){\line(0,1){24}}\put(26,24){\line(1,0){24}}
\put(22,-30){\small $\nu^2$}
\end{picture}}
\put(140,50){\begin{picture}(50,50)
\linethickness{0.1mm}\put(0,25){\line(1,0){50}}
\linethickness{0.5mm}\put(25,0){\line(0,1){50}}
\put(22,-30){\small $\lambda q^{\Nop_j}$}
\end{picture}}
\put(210,50){\begin{picture}(50,50)
\linethickness{0.5mm}\put(0,25){\line(1,0){50}}
\linethickness{0.1mm}\put(25,0){\line(0,1){50}}
\put(22,-30){\small $\mu q^{\Nop_j}$}
\end{picture}}
\put(280,50){\begin{picture}(50,50)
\linethickness{0.1mm}\put(0,25){\line(1,0){25}}\put(25,25){\line(0,1){25}}
\linethickness{0.5mm}\put(25,0){\line(0,1){25}}\put(25,25){\line(1,0){25}}
\put(22,-30){\small $\nu\xop_j$}
\end{picture}}
\put(350,50){\begin{picture}(50,50)
\linethickness{0.5mm}\put(0,25){\line(1,0){25}}\put(25,25){\line(0,1){25}}
\linethickness{0.1mm}\put(25,0){\line(0,1){25}}\put(25,25){\line(1,0){25}}
\put(20,-30){\small $\nu\yop_j$}
\end{picture}}
\end{picture}
\caption{The ``weights'' of $j^{\;\textrm{th}}$ vertex,
$j=1,...,NM$.} \label{fig-weights}
\end{figure}
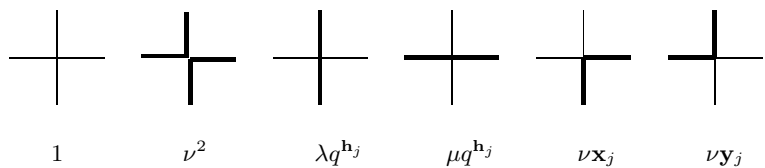

\noindent Here $\lambda,\mu,\nu$ ($\nu^2\stackrel{\textrm{def}}{=}
-q^{-1}\lambda\mu$) are numeric parameters common for whole
lattice. The key point is that we will regard the inhomogeneous
``weights'' $q^{\Nop_j},\xop_j,\yop_j$ as generators of local
$q$-oscillator algebra $\alg$,
\begin{equation}\label{q-osc}
\alg\;:\ \ \xop\yop=1-q^{2+2\Nop}\;,\;\;\;
\yop\xop=1-q^{2\Nop}\;,\;\;\; \xop q^{\Nop}= q^{\Nop+1}
\xop\;,\;\;\; \yop q^{\Nop}= q^{\Nop-1} \yop\;.
\end{equation}
The vertex index $j$ in Fig.\ref{fig-weights} stands for the
$j^{\;\textrm{th}}$ component of the tensor power $\alg^{\otimes
NM}$. In this paper we will imply the Fock space representation of
$q$-oscillator ($\mathrm{Spectrum}(\Nop)=0,1,2,\dots$). In the
limit $q\to 1$ the model becomes the completely inhomogeneous
free-fermion six-vertex model \cite{Korepanov}.

Partition function $Z$ is a polynomial of two parameters $\lambda$
and $\mu$ (recall, $\nu^2=-q^{-1}\lambda\mu$), its operator-valued
coefficients belong to $\alg^{\otimes NM}$. The main result of
this letter is the commutativity of $Z$,
\begin{equation}\label{commutativity}
Z(\lambda,\mu) \; Z(\lambda',\mu') \;=\; Z(\lambda',\mu') \;
Z(\lambda,\mu)\;\;\;\;\forall\;\lambda,\mu,\lambda',\mu'\;.
\end{equation}
Therefore, $Z(\lambda,\mu)$ is the layer-to-layer transfer matrix
of the quantum mechanical model in wholly discrete $2+1$
dimensional space-time. This is the simplified formulation of
\emph{auxiliary} $q$-oscillator lattice \cite{First}.

\begin{figure}[ht]
\setlength{\unitlength}{0.20mm}
\begin{picture}(200,200)
\thinlines
\put(25,0){\line(0,1){200}} \put(75,0){\line(0,1){200}}
\put(125,0){\line(0,1){200}} \put(175,0){\line(0,1){200}}
\put(0,25){\line(1,0){200}} \put(0,75){\line(1,0){200}}
\put(0,125){\line(1,0){200}} \put(0,175){\line(1,0){200}}
\put(30,0){\scriptsize $\alpha_1$} \put(80,0){\scriptsize
$\alpha_2$} \put(130,0){\scriptsize $\alpha_3$}
\put(155,0){$\dots$} \put(180,0){\scriptsize $\alpha_{N}$}
\put(0,30){\scriptsize $\beta_1$} \put(0,150){$\vdots$}
\put(0,80){\scriptsize $\beta_2$} \put(0,130){\scriptsize
$\beta_3$} \put(0,180){\scriptsize $\beta_{M}$}
%
\thicklines
\put(20,20){\line(1,1){10}}\put(20,70){\line(1,1){10}}
\put(20,120){\line(1,1){10}}\put(20,170){\line(1,1){10}}
\put(70,20){\line(1,1){10}}\put(70,70){\line(1,1){10}}
\put(70,120){\line(1,1){10}}\put(70,170){\line(1,1){10}}
\put(120,20){\line(1,1){10}}\put(120,70){\line(1,1){10}}
\put(120,120){\line(1,1){10}}\put(120,170){\line(1,1){10}}
\put(170,20){\line(1,1){10}}\put(170,70){\line(1,1){10}}
\put(170,120){\line(1,1){10}}\put(170,170){\line(1,1){10}}
\end{picture}
\caption{Labeling of the lines of the $q$-oscillator lattice (a
layer-to-layer transfer matrix).}\label{fig-lattice}
\end{figure}
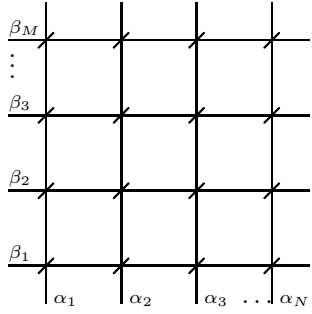

Turn to the proof of Eq.(\ref{commutativity}). It is convenient to
combine the weights of Fig.\ref{fig-weights} into a
six-vertex-type matrix $L_{\alpha\beta}$ acting in the product of
two dimensional vector spaces $V_\alpha\otimes V_\beta$, $V\equiv
\mathbb{C}^2$:
\begin{equation}\label{quantum-L}
L_{\alpha\beta}(\alg_j;\lambda,\mu)\;=\; \left(\begin{array}{cccc}
1 & 0 & 0
& 0\\
0 & \lambda q^{\Nop_j} & \nu \yop_j & 0 \\ 0 & \nu \xop_j & \mu q^{\Nop_j} & 0 \\
0 & 0 & 0 & \nu^2
\end{array}\right)\;,\;\;\;\nu^2=-q^{-1}\lambda\mu\;.
\end{equation}
Let the lines of the lattice are labeled by the indices $\alpha_n$
and $\beta_m$, $1\leq n\leq N$ and $1\leq m\leq M$ as it is shown
in Fig.\ref{fig-lattice}. The vertices of the lattice are labeled
by $j=(n,m)$.

The ``partition function'' $Z$ may be written as
\begin{equation}\label{pf-as-trace}
Z(\lambda,\mu)\;=\;\mathop{\textrm{Trace}}_{V_{\boldsymbol{\alpha}}\otimes
V_{\boldsymbol{\beta}}}\biggl(
\mathbf{L}_{\boldsymbol{\alpha}\boldsymbol{\beta}}(\lambda,\mu)
\biggr)\;,\;\;\;\;
\mathbf{L}_{\boldsymbol{\alpha}\boldsymbol{\beta}}(\lambda,\mu)\;=\;
\prod_{n}^{\curvearrowleft}\prod_m^\curvearrowleft\;
L_{\alpha_n\beta_m}(\alg_{n,m};\lambda,\mu)\;,
\end{equation}
where $\ds \prod^\curvearrowleft_n f_n=f_N f_{N-1} \dots f_2 f_1$,
$\ds \prod^\curvearrowleft_m f_m=f_M f_{M-1} \dots f_2 f_1$, $\ds
V_{\boldsymbol{\alpha}}=\mathop{\otimes}_{n=1}^N V_{\alpha_n}$ and
$\ds V_{\boldsymbol{\beta}}=\mathop{\otimes}_{m=1}^M V_{\beta_m}$.

It is known \cite{BS-TE}, the commutativity of layer-to-layer
transfer matrices follows from a tetrahedron equation. In
particular, the commutativity (\ref{commutativity}) follows from
\begin{equation}\label{LTE}
\begin{array}{l}
\ds M_{\alpha\alpha'}\left(\alg_0;\frac{\mu}{\mu'}\right) \;
M_{\beta\beta'}\left(\alg_0;\frac{\lambda'}{\lambda}\right) \;
L_{\alpha\beta}(\alg;\lambda,\mu) \; L_{\alpha'\beta'}(\alg;\lambda',\mu')\;=\\
\\
\ds  L_{\alpha'\beta'}(\alg;\lambda',\mu') \;
L_{\alpha\beta}(\alg;\lambda,\mu) \;
M_{\beta\beta'}\left(\alg_0;\frac{\lambda'}{\lambda}\right) \;
M_{\alpha\alpha'}\left(\alg_0;\frac{\mu}{\mu'}\right)\;,
\end{array}
\end{equation}
where matrix elements of $M(\alg_0;\xi)$ belong to an additional
copy $\alg_0$ of $q$-oscillator:
\begin{equation}\label{M}
M(\alg_0;\xi)\;=\;\left(\begin{array}{cccc} \xi^{\Nop_0} & 0 & 0 &
0 \\ 0 & \lambda_0 (-q\xi)^{\Nop_0} & \nu_0
\xi^{-1/2+\Nop_0}\yop_0
& 0 \\
0 & \nu_0 \xi^{1/2+\Nop_0}\xop_0 & \mu_0 (-q\xi)^{\Nop_0} & 0 \\
0 & 0 & 0 & \nu_0^2\xi^{\Nop_0}\end{array}\right)\;,\;\;\;
\nu_0^2= q^{-1}\lambda_0\nu_0\;.
\end{equation}
Equation (\ref{LTE}) may be verified directly (in auxiliary spaces
$V_\alpha\otimes\cdots\otimes V_{\beta'}$, it is just $16\times
16$ matrix equation), the commutativity (\ref{commutativity}) may
derived by the repeated use of (\ref{LTE}) for the forms
(\ref{pf-as-trace}).

The layer-to-layer transfer matrix $Z(\lambda,\mu)$ may be
interpreted in the terms of the two-dimensional quantum inverse
scattering method and quantum groups (\cite{First} and e.g.
\cite{SFT,RTF}). Transfer matrix (\ref{pf-as-trace}) is the $2d$
transfer matrix $\ds
Z(\lambda,\mu)=\mathop{\textrm{Trace}}_{V_{\boldsymbol{\beta}}}\prod^\curvearrowleft_n
\mathcal{L}_{\boldsymbol{\beta}}^{(n)}(\lambda,\mu)$ for the
length-$N$ chain of the Lax operators
\begin{equation}\label{Lax}
\mathcal{L}_{\boldsymbol{\beta}}^{(n)}(\lambda,\mu)\;=\;
\mathop{\textrm{Trace}}_{V_{\alpha_n}} \prod^\curvearrowleft_m
L_{\alpha_n,\beta_m}(\alg_{n,m};\lambda,\mu)\;,
\end{equation}
Due to the six-vertex structure of (\ref{quantum-L}), the Lax
operator (\ref{Lax}) has the block-diagonal form (in the
combinatorial formulation of Fig.\ref{fig-weights} it means the
conservation of the number of bold edges on the left and right of
$\alpha_n^{\;\textrm{th}}$ column of Fig.\ref{fig-lattice}):
\begin{equation}\label{decomposition}
\mathcal{L}_{\boldsymbol{\beta}}(\lambda,\mu)\;=\;
\mathop{\textrm{\Large $\oplus$}}_{m=0}^{M} (-q^{-1}\mu)^m\;
L_{\omega_m}(\lambda)
\end{equation}
where $L_{\omega_m}$ is the Lax operator for $m^{\;\textrm{th}}$
fundamental representation $\pi_{\omega_m}$ of
$\mathcal{U}_q(\widehat{sl}_M)$ in the auxiliary space\footnote{It
is implied, $\omega_m$ are the dominant weights of
$\mathcal{U}_q(\widehat{sl}_M)$, $\pi_{\omega_0}$ and
$\pi_{\omega_M}$ stand for scalar representations.}. In the
quantum space, the Lax operator (\ref{Lax}) acts in $F^{\otimes
M}$, where $F$ is a representation space of $q$-oscillator $\alg$.
The Lax operator (\ref{Lax}) has the central element
\begin{equation}\label{J-center}
q^{J_n}\;=\;q^{\Nop_{n,1}+\Nop_{n,2}+\cdots +\Nop_{n,M}}\;.
\end{equation}
For the Fock space representation of $q$-oscillators,  $\ds
F^{\otimes M}=\mathop{\oplus}_{J=0}^\infty \pi_{J\omega_1}$ is the
direct sum of rank-$J$ symmetrical tensor representations of
$\mathcal{U}_q(\widehat{sl}_M)$. The $R$-matrix for the Lax
operators (\ref{Lax}) follows from (\ref{LTE}):
\begin{equation}
R_{\boldsymbol{\beta},\;\boldsymbol{\beta}'}
\left(\frac{\lambda}{\lambda'}\right) \;=\;
\mathop{\textrm{Trace}}_{\alg_0} \prod^\curvearrowleft_m
M_{\beta_m^{},\;\beta_m'}\left(\alg_0;\frac{\lambda'}{\lambda}\right)\;.
\end{equation}
Matrix elements of $M$ (\ref{M}) depend on two extra parameters
$\lambda_0,\mu_0$. They produce the decomposition
\begin{equation}
R_{\boldsymbol{\beta},\;\boldsymbol{\beta}'}
\left(\frac{\lambda}{\lambda'}\right) \;=\; \mathop{\textrm{\Large
$\oplus$}}_{m,m'=0}^M \; \lambda_0^m \mu_0^{m'}
R_{\omega_{m^{}},\,\;\omega_{m'}}
\left(\frac{\lambda}{\lambda'}\right)
\end{equation}
where $R_{\omega_{m^{}},\;\omega_{m'}}$ is the $R$-matrix for
$\pi_{\omega_{m^{}}}\otimes \pi_{\omega_{m'}}$ fundamental
representations of $\mathcal{U}_q(\widehat{sl}_M)$.

The definition of the ``partition function'' $Z$ was initially
$N\leftrightarrow M$ invariant. In particular, the alternative to
(\ref{Lax}) product $\ds\mathop{\textrm{Trace}}_{V_\beta}
\prod^\curvearrowleft_n L_{\alpha_n,\beta}(\alg_n;\lambda,\mu)$ is
the Lax operator for $\mathcal{U}_q(\widehat{sl}_N)$ with the
spectral parameter $\mu$, while $M$ becomes the length of
$\mathcal{U}_q(\widehat{sl}_N)$ chain. Central elements of
$\mathcal{U}_q(\widehat{sl}_N)$ $L$-operators are (cf.
(\ref{J-center}))
\begin{equation}
q^{K_m}\;=\; q^{\Nop_{1,m}+\Nop_{2,m}+\cdots + \Nop_{N,m}}\;.
\end{equation}
Since both sets of occupation numbers $J_n$ and $K_m$ are
integrals of motion, their eigenvalues define a sub-sector of the
$q$-oscillator model. For example, if $M=2$ and the lattice is
interpreted as the $\mathcal{U}_q(\widehat{sl}_2)$ chain with the
length $N$, the choice $J_n=1$ gives us the six-vertex model (in
general, $\pi_{J\omega_1}$ is spin $J/2$ representation of
$\mathcal{U}_q(sl_2)$), whereas $K_1$ and $K_2$ stand for the
numbers of spins up and spins down.

We would like to conclude the letter by the announcement of the
universal form of the nested Bethe Ansatz equation for $Z$. The
explicit polynomial decomposition of $Z$ is
\begin{equation}\label{polynomial}
Z(\lambda,\mu) = \sum_{n=0}^N \sum_{m=0}^{M} \lambda^{M n} \mu^{N
m} z_{n,m}\;\equiv\; \sum_{m=0}^M \mu^{Nm}
T_{\omega_m}^{(sl_M)}(\lambda^M)\;\equiv\; \sum_{n=0}^N
\lambda^{Mn} T_{\omega_n}^{(sl_N)}(\mu^N)\;.
\end{equation}
Let $\uop,\vop$ be an additional auxiliary Weyl pair, $\ds
\uop\vop=q^2\vop\uop$, serving the following notations:
\begin{equation}\label{Q-rules}
\langle Q|u\rangle\;=\;Q(u)\;,\;\;\; \langle Q|\uop|u\rangle \;=\;
uQ(u)\;,\;\;\; \langle Q|\vop|u\rangle\;=\;v Q(q^2u)\;.
\end{equation}
Let now (cf. (\ref{polynomial}), where the last but one expression
is related to $\mathcal{U}_q(\widehat{sl}_M)$ Bethe Ansatz)
\begin{equation}
J(\uop,\vop)\;=\;\sum_{n=0}^N \sum_{m=0}^{M} (-q)^{-nm} \uop^{n}
\vop^m z_{n,m}\;\equiv\; \sum_{m=0}^M \vop^m
T_{\omega_m}^{(sl_M)}\biggl((-q)^m\uop\biggr)\;.
\end{equation}
Then the nested Bethe Ansatz equation for
$\mathcal{U}_q(\widehat{sl}_M)$ chain (notations (\ref{Q-rules})
are taken into account) is
\begin{equation}\label{TQ}
\langle Q|\;J(\uop,\vop)\; |u\rangle\;\equiv\; \sum_{m=0}^M v^m
Q(q^{2m}u) T_{\omega_m}^{(sl_M)}\biggl( (-q)^m u\biggr)\; =0\;.
\end{equation}
Extra condition $Q(0)=1$ is an $M^{\;\textrm{th}}$ power equation
for $v$ entering the definition (\ref{Q-rules}), its
$m^{\;\textrm{th}}$ solution $v=v_m$ corresponds to an order $K_m$
polynomial $Q=Q_m(u)$. The set of $[v_m,Q_m(u)]_{m=1..M}$ is a
complete set of independent solutions of $M^{\;\textrm{th}}$ order
linear equation (\ref{TQ}). In the same way and with the same
results one may consider other realizations of Weyl algebra, for
instance equation $\langle v| J(\uop,\vop) |\overline{Q}\rangle=0$
for polynomials $\langle v|\overline{Q}\rangle=\overline{Q}(v)$
and the last expression in (\ref{polynomial}) gives the dual
$\mathcal{U}_q(\widehat{sl}_N)$ Bethe Ansatz\footnote{Objects
$\langle u|\overline{Q}\rangle$ and $\langle Q|v\rangle$ are
polynomials for anti-Fock space representation
$\mathrm{Spectrum}(\Nop)=-1,-2,-3,\dots$.}.

In this letter we considered rectangular lattice with homogeneous
$\lambda,\mu$. The results of this paper may be generalized to the
case of a lattice of any shape with inhomogeneous set of
$\lambda_j,\mu_j$. This, as well as the $3d$-invariant derivation
of (\ref{TQ}), is the subject of forthcoming papers.

\noindent\textbf{Acknowledgments} The author would thank R.Baxter,
M.Batchelor, V.Bazhanov and V.Mangazeev for valuable discussions.


\begin{thebibliography}{*}

\bibitem{Baxter-book}
R. J. Baxter, ``Exactly Solved Models in Statistical Mechanics'',
Academic Press, London, 1972


\bibitem{Korepanov}
I. G. Korepanov, ``Algebraic integrable dynamical systems, $2+1$
dimensional models on wholly discrete space-time, and
inhomogeneous models of 2-dimensional statistical physics'',
\emph{Preprint} solv-int/9506003 (1996)

\bibitem{First}
V. V. Bazhanov and S. M. Sergeev, ``Zamolodchikov's tetrahedron
equation and hidden structure of quantum groups'', \emph{preprint}
arXiv:hep-th/0509181 (2005)

\bibitem{BS-TE}
V. V. Bazhanov and Yu. G. Stroganov, ``Conditions of commutativity
of transfer-matrices on a multidimensional lattice'', \emph{Theor.
Math. Phys.} \textbf{52} (1982) 685-691

\bibitem{SFT}
E. K. Skljanin, L. A. Takhtadzhyan and L. D. Faddeev, ``Quantum
inverse problem method. I.'', \emph{Theor. Math. Phys.}
\textbf{40} (1979) 688-706

\bibitem{RTF}
N. Yu. Reshetikhin, L. A. Takhtadzhyan and L. D.  Faddeev,
``Quantization of Lie groups and Lie algebras'', \emph{Leningrad
Math. J.} \textbf{1} (1990) 193--225



\end{thebibliography}
\end{document}